\documentclass[a4paper,11pt]{article}
\usepackage{pos}

\title{Search for the Axion-Like-Particles in the $\eta\to\pi^{+}\pi^{-}e^{+}e^{-}$ decay with HADES detector}

\author*[a]{Marcin Zieliński}
\author[a,b]{Krzysztof Prościński}
\author[a]{Piotr Salabura}
\onbehalf{\\for the HADES Collaboration}

\affiliation[a]{M. Smoluchowski Institute of Physics, Jagiellonian University, 30-348 Kraków, Poland}
\affiliation[b]{Doctoral School of Exact and Natural Sciences, Jagiellonian University, 31-007 Kraków, Poland}

\emailAdd{marcin.zielinski@uj.edu.pl}

\abstract{The dark matter and existence of new particles are now a possible explanation of several physics phenomena which evade the predictions of the Standard Model. In this context Axion-Like-Particles (ALP) with masses in the MeV to GeV range with additional Peccei-Quinn breaking contribution, and which are coupled to the Standard Model have been postulated. To search for the existence of such new particles, we have launched dedicated analysis of a high statistics data sample collected by High-Acceptance Di-Electron Spectrometer (HADES) operating at GSI in Darmstadt. In particular, we study $\eta$ meson decays into  $\pi^{+}\pi^{-}e^{+}e^{-}$, where hypothesized isoscalar gauge boson $a$ could be produced in the intermediate state $\eta\to\pi^{+}\pi^{-}a$ decaying predominantly to $e^{+}e^{-}$. In this report we describe the analysis strategy we applied to search for a resonant peak in the dilepton invariant mass spectrum $\eta\to\pi^{+}\pi^{-}a\to \pi^{+}\pi^{-}e^{+}e^{-}$ and present the method for event selection and particle identification.}

\FullConference{The 21st International Conference on Hadron Spectroscopy and Structure (HADRON2025)\\
 27 - 31 March, 2025\\
Osaka University, Japan\\}

\begin{document}
\maketitle

\section{Introduction}
The Standard Model (SM) is now a well known and established theory describing the
particles and their strong, electromagnetic, and weak interactions. A major contribution
to this field was made by the ATLAS and CMS experiments conducted at the Large Hadron Collider
(LHC) when the Higgs boson was discovered~\cite{ATLAS:2012yve,CMS:2012qbp}. This groundbreaking finding confirms the high accuracy of the Standard Model predictions. However, many observed phenomena are still not well understood, which indicates that the SM is not a complete theory providing a full description of all interactions in the Universe. 
Particularly, it fails when trying to explain, e.g., matter and antimatter asymmetry in the Universe, the acceleration of the Universe expansion, and neutrino masses. Additionally, there is compelling evidence that physics Beyond the Standard Model (BSM) may involve new particles or force mediators that violate certain discrete symmetries, particularly CP symmetry. 
The most suitable strategy to search for new phenomena is through hidden dark sectors~\cite{Batell:2009di,Batell:2009yf}, characterized by light states below the electroweak scale that couple only very weakly to the Standard Model. It can be experimentally accessible via particles in the MeV–GeV mass range, which are coupled to the Standard Model. 
To this end, one promising option is a pseudoscalar sector where one can expect leptonic, radiative, and hadronic decays of new particles such as QCD axions, dark photons, or other axion-like particles (ALPs). Originally, the existence of an axion particle was proposed by Peccei-Quinn (PQ) as a Two-Higgs-Doublet Model with a common breaking mechanism for the electroweak and PQ symmetries. Later on, the theory was further developed by Wilczek~\cite{PhysRevLett.40.279}. He  included constraints from the low energy QCD regime and postulated the existence of a QCD axion. However, until now, its existence has been excluded by searches in hadronic decays and beam dump experiments~\cite{Bjorken:1988as,Bross:1989mp,Blumlein:1990ay,Riordan:1987aw}, and the search program was almost abandoned in the late 1980s. 

However, recently, a new set of calculations was performed that restored the QCD axion with a mass $m_a = \mathcal{O}(1 - 100)$~MeV and decay constants $f_a = \mathcal{O}(1 - 10)$~GeV~\cite{Alves:2017avw}, together with a more general type of new Axion-Like Particles (ALP) that includes additional PQ-breaking contributions to their masses~\cite{Alves:2020xhf}.
Another hint of a possible new particle was claimed by the ATOMKI collaboration~\cite{Krasznahorkay:2015iga}, which has measured the emission of $e^{+}e^{-}$ pairs in several transitions of the $^8Be$ isotope. In these processes, a "bumplike" excess in the invariant mass and opening angle distributions of $e^+ e^-$ pairs has been observed and attributed to a new particle with a mass around 17~MeV ($X_{17}$). One of the first explanations for this observation was based on a protophobic gauge boson~\cite{Feng:2016ysn}. Later, another explanation postulated a piophobic particle that has suppressed mixing with the neutral pion~\cite{Alves:2017avw}.
This discovery has triggered searches in particle decays, where the hadronic couplings of the new piophobic ALP could lead to stricter constraints. In this approach, one of the promising processes is rare and very rare $\eta$ and $\eta^{\prime}$ meson decays, which are suppressed by the SM. To this end, the considered axio-hadronic decays are three body final states $\eta\to\pi^{+}\pi^{-}a$ and $\eta\to\pi^{0}\pi^{0}a$~\cite{Alves:2017avw}.
The leading-order potential term in the chiral Lagrangian directly contributes to the amplitudes of these processes, which could lead to sizable branching ratios. Moreover, assuming that the $a$ is short-lived, it predominantly decays to $e^+ e^-$ ($BR\approx 1)$, which also provides a solution to avoid limits from beam dump and fixed target experiments. By using the Resonance Chiral Theory ($\chi$RT), which encodes the most prominent features of non perturbative strong dynamics~\cite{Ecker:1988te,Ecker:1989yg} and incorporates low-lying resonances~\cite{Oller:1997ti}, the rates for $\eta$/$\eta^{\prime}$ axio-hadronic processes were calculated~\cite{Alves:2017avw}. Estimations based on the $\chi$RT predicted branching ratios spanning two orders of magnitude $BR(\eta\to\pi\pi a) \approx (10^{-4} - 10^{-2})$. More recent studies, including final state pion rescattering effects, estimate a  lower upper limit in the range of $10^{-4} - 10^{-6}$~\cite{Alves:2024dpa}.

Experimentally, to probe the possible existence of the $a$ particle, one can use the $\eta \rightarrow \pi^{+}\pi^{-}e^{+}e^{-}$ final state. This decay was already observed by several experiments and studied in view of possible CP violation in flavor-conserving reactions; however, it lacked sensitivity at such low masses ~\cite{KLOE:2008arm,Adlarson:2015zta}.
Most recently, this decay was measured by the BESIII experiment, which also implies no CP-violation with the collected statistical sample~\cite{BESIII:2025cky}. In addition, within the BESIII studies, the limits for the possible existence of ALP in the process $\eta\to\pi^{+}\pi^{-}a$ in the mass range of 0 - 200~MeV/c\(^{2}\) were established at the level of $BR=(3.0 - 53.0)\times10^{-3}$ with a significance of about 0.5~$\sigma$. Similar studies are foreseen in future experiments such as REDTOP~\cite{Zielinski:2025wfa}, SHiP~\cite{Alekhin:2015byh}, and possibly also at the JEF facility~\cite{Somov:2024jiy}. 

In order to study these effects, a dedicated analysis of high statistics data from proton-proton interactions at a kinetic beam energy of 4.5 GeV collected by the HADES detector~\cite{HADES:2009aat} has been launched. A data sample with an integrated luminosity of approximately 5.5~pb\(^{-1}\) was analyzed ~\cite{HADES:2020pcx}. In this article, we will describe the basic selection steps that are crucial to identify the final state particles $\pi^{+}\pi^{-}e^{+}e^{-}$ from $\eta$ decay. 

\section{Identification of the $\eta\rightarrow \pi^{+}\pi^{-}e^{+}e^{-}$ decay}
To identify events that constitute the possible signal for the $\eta \rightarrow \pi^{+}\pi^{-}e^{+}e^{-}$ decay, individual particles are selected as a first step. Initially, all particle candidates from one event are divided into general categories: positive or negative and lepton or hadron. 
The latter is achieved based on the signals registered in the RICH detector, which is hadron blind and has a high efficiency for detecting leptons~\cite{Fortsch:2020rjg,Becker:2023arw}. The considered event categories are shown in Tab.~\ref{tab:tab1}.
\begin{table}[!h]
        \centering
        \begin{tabular}{|c|c|c|}
            \hline
            {\bf Charge}    &{\bf RICH signal}  &{\bf Category}\\\hline\hline
            positive        &present            &lepton (+)\\ \hline
            negative        &present            &lepton (--)\\ \hline
            positive        &not present        &hadron (+)\\ \hline
            negative        &not present        &hadron (--)\\ \hline
        \end{tabular}
        \caption{Initial categories for selection of particles based on the information delivered by RICH detector.}
        \label{tab:tab1}
    \end{table}

\begin{figure}[ht]
  \centerline{
  \includegraphics[width=6.25cm]{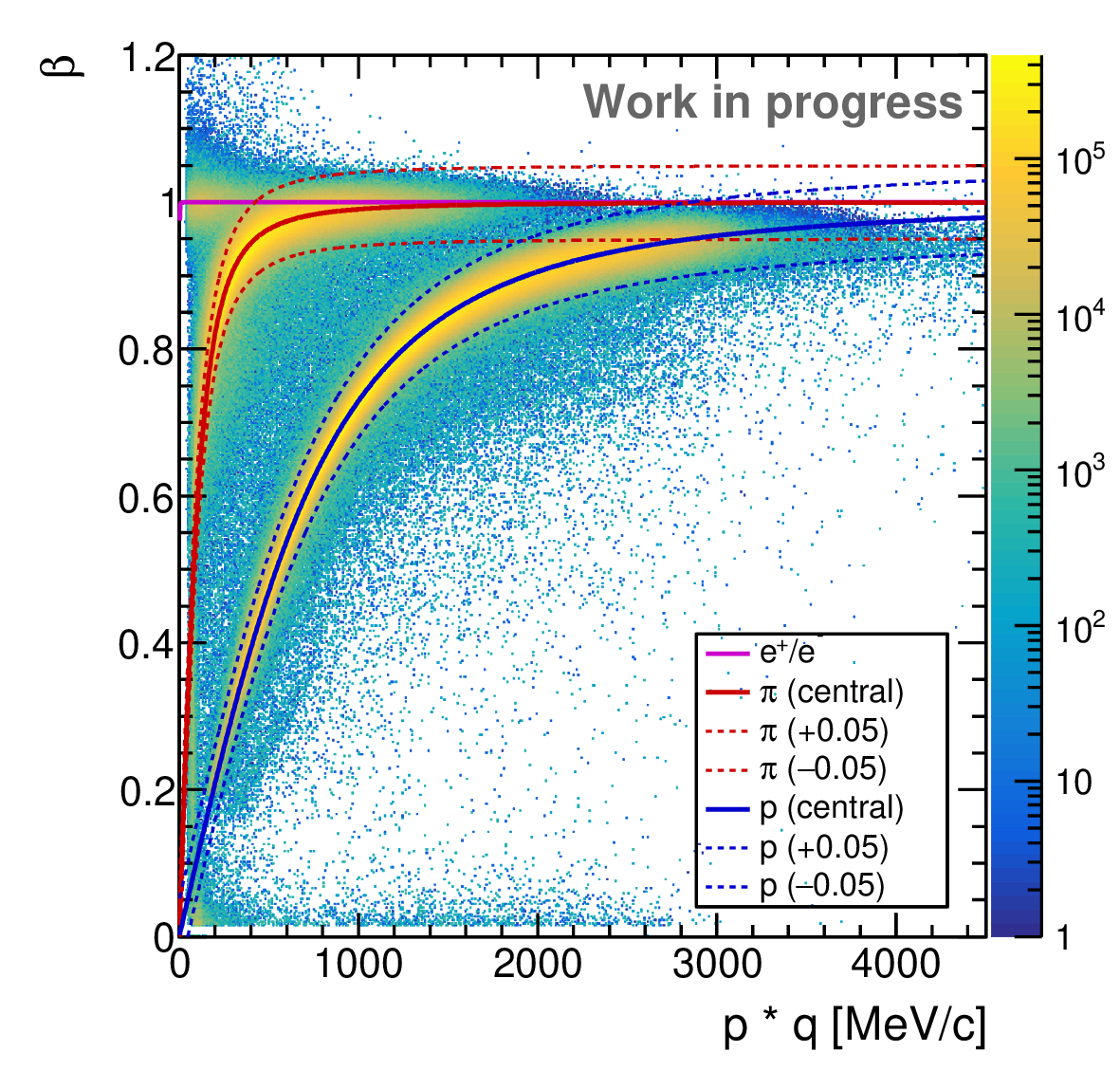}
  \includegraphics[width=6.25cm]{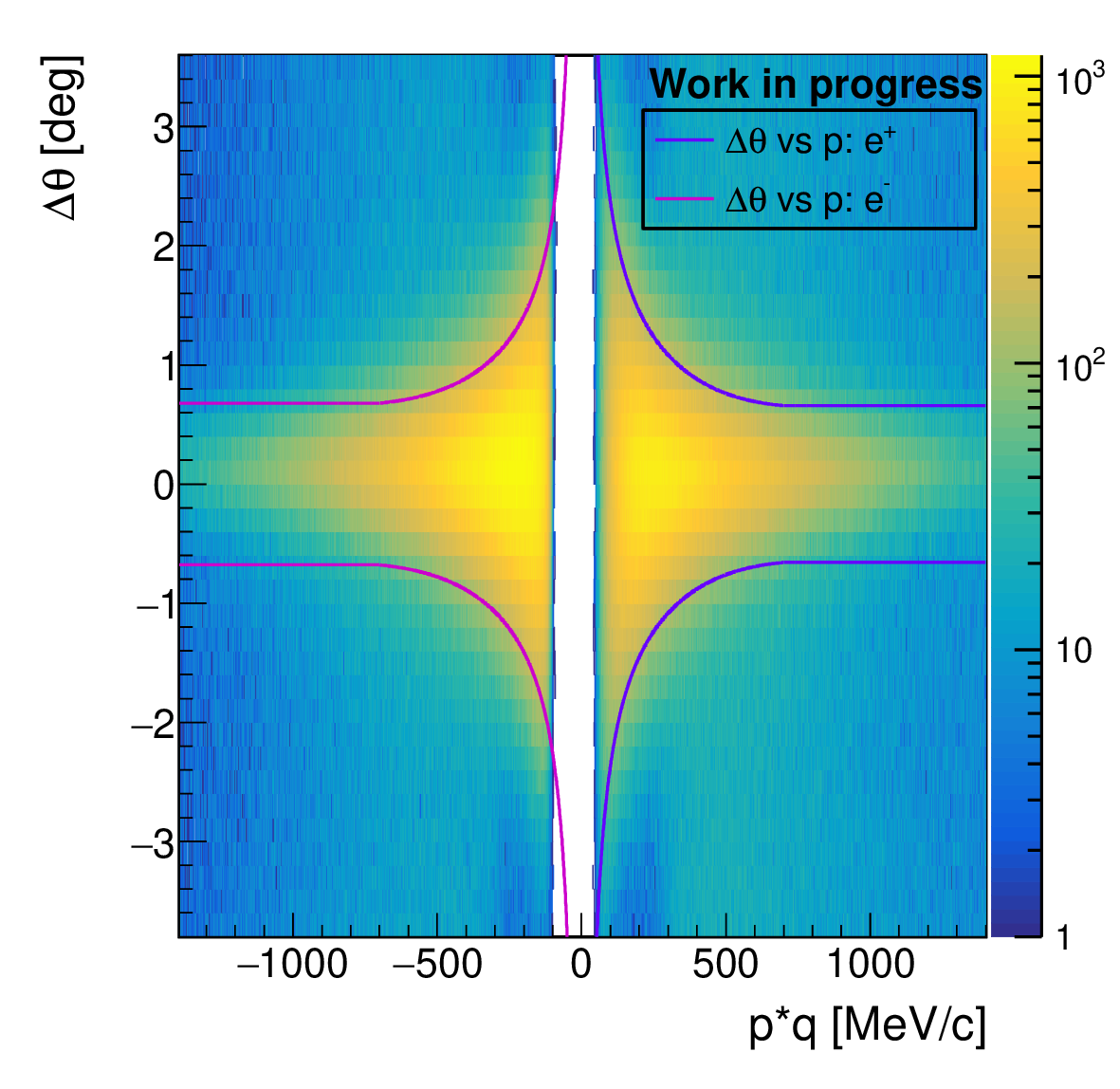}
  }
\caption{Base criteria for particle selection and identification. 
{\bf (left)} Distribution of particle velocity $\beta$ as a dependence of momentum $p\cdot q$. The superimposed red and blue lines indicate the selection cuts for pions and protons, respectively.  
{\bf (right)} Example of the angular distribution of $\Delta\theta$ as a function of momentum for leptons. The selection cuts are indicated by superimposed lines: purple for electrons and magenta for positrons.}
\label{Fig:f1}
\end{figure}

After initial identification, additional selection criteria are applied to properly distinguish between charged leptons and pions. 
This selection is performed  by applying a graphical cut in the $\beta$ vs. momentum distribution, as shown in Fig.~\ref{Fig:f1}~(left). The cut  was  established based on the Monte Carlo simulation of the HADES detector.  Furthermore, leptons are identified by parameterizing a two dimensional dependence of the differences in the polar ($\Delta \theta = \theta_{RICH} - \theta_{MDC}$) and azimuthal ($\Delta \phi = \phi_{RICH} - \phi_{MDC}$) angles of tracks measured in the MDC and RICH detectors and the particle momentum. The respective  exemplary distribution for the $\Delta \theta$ for leptons is shown in Fig.~\ref{Fig:f1}~(right).

Next, in order to select the candidates for the searched decay, an event hypothesis that includes all required particles must be defined. In this analysis, we require that each event has at least properly identified $\pi^+$, $\pi^-$, $e^+$, and $e^-$, satisfying the selection criteria described above. Moreover, in addition to the main hypothesis, two additional hypotheses ($\pi^+$, $\pi^-$, $e^+$, $e^+$) and ($\pi^+$, $\pi^-$, $e^-$, $e^-$) must be analyzed to properly calculate the lepton combinatorial background.  

\begin{figure}[b]
\centerline{%
\includegraphics[width=6.25cm]{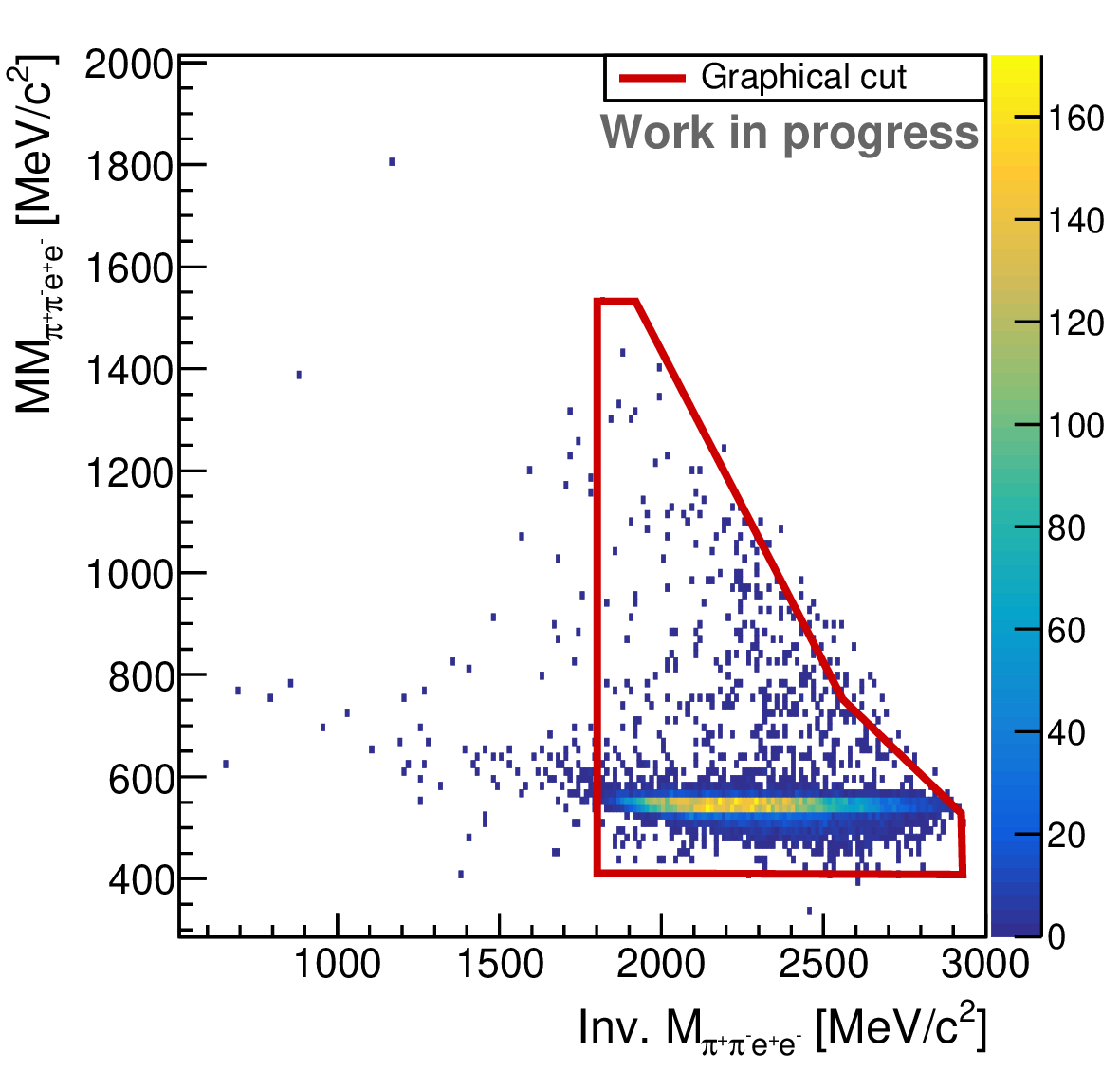}
\includegraphics[width=6.25cm]{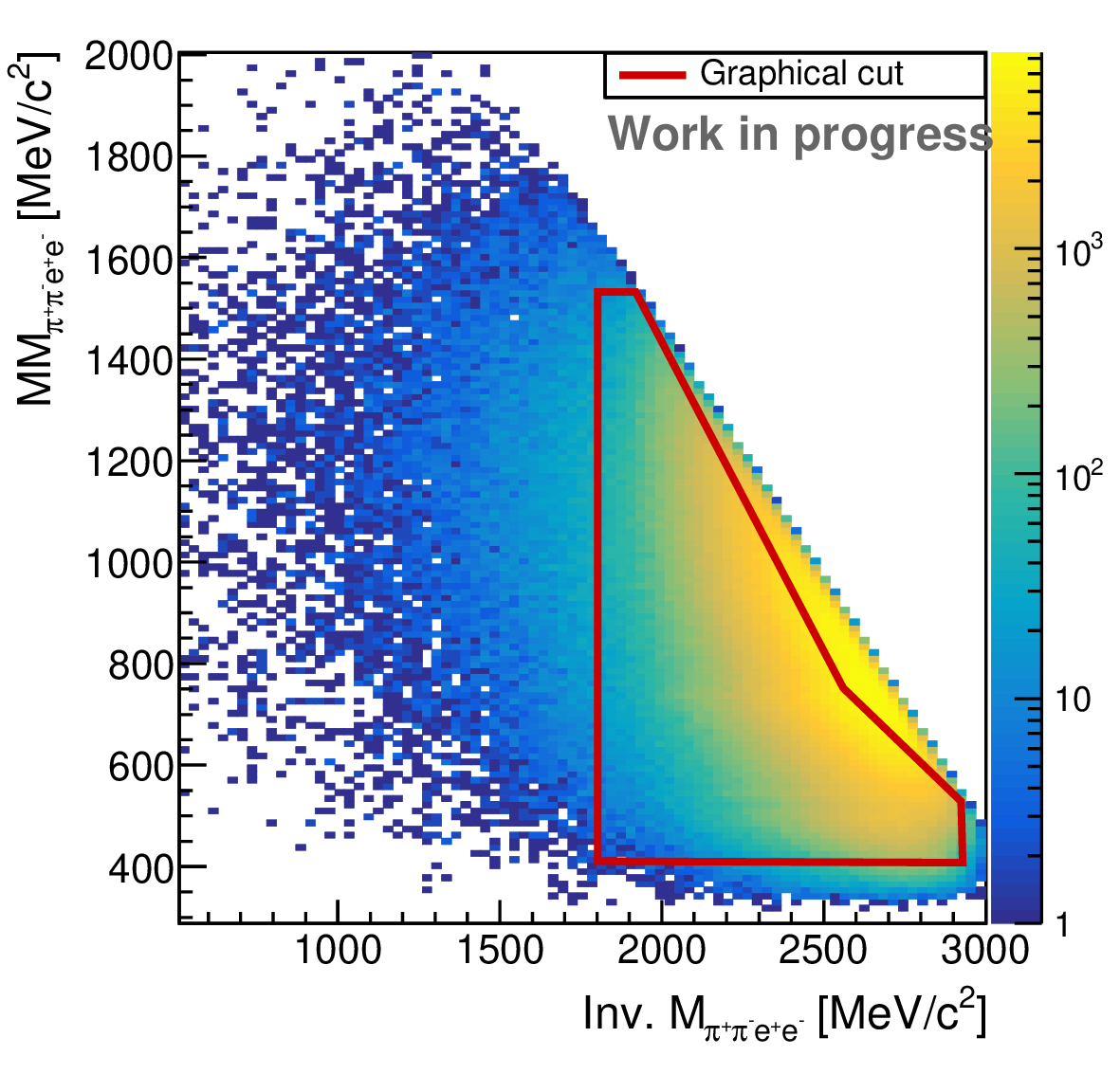}
}
\caption{
The dependence of the missing mass and invariant mass for all reconstructed and identified particles fulfilling the hypothesis of $\pi^+ \pi^- e^+ e^-$: {\bf (left)} for the Monte Carlo simulation of the  $\eta \to \pi^+\pi^- e^+ e^-$ decay, and {\bf (right)} for the experimental data. Superimposed red line on both panels indicates the chosen constrain.}
\label{Fig:f2}
\end{figure}
The investigated decay $\eta\rightarrow \pi^{+}\pi^{-}e^{+}e^{-}$ is a rare process with a branching fraction of BR~=~(2.68~$\pm$~0.11)$\times10^{-4}$~\cite{ParticleDataGroup:2024cfk}. Moreover, the identification of light mesons such as $\eta,\eta^{\prime}$ in proton-proton collisions at higher energies is affected by significant hadronic background contamination. In particular, the multipion background, e.g., $2\pi,3\pi,....$, dominates all invariant mass distributions used for signal identification~\cite{Zielinski:2011dt}. Therefore, to suppress background, a series of consecutive constraints has been implemented in the data sample, as follows:
(i) $z$ coordinate of the reconstructed vertex in the range of -200~mm to 0~mm,
(ii) correlation of $\pi^+ \pi^- e^+ e^-$ missing versus invariant masses in a restricted range, as shown in Fig.~\ref{Fig:f2}~(right),
(iii) opening angle for the ($e^+e^-$)($\pi^+\pi^-$) less than 50\textdegree,
(iv) invariant mass of $\pi^+\pi^-$ less than 480~MeV,
(v) opening angle in the center of mass frame for the ($e^+e^-$)($\pi^+\pi^-$) greater than 140\textdegree. The above conditions were derived from Monte Carlo simulations  of the signal and background, utilizing apparent differences in the respective distributions.
As an example of such distributions, the correlation of the $\pi^+\pi^-e^+e^-$ missing mass and invariant mass is shown in Fig.~\ref{Fig:f2} for the MC simulations of the signal (left) and for the experimental data (right).

\begin{figure}[tb]
  \centerline{
   \includegraphics[width=6.25cm]{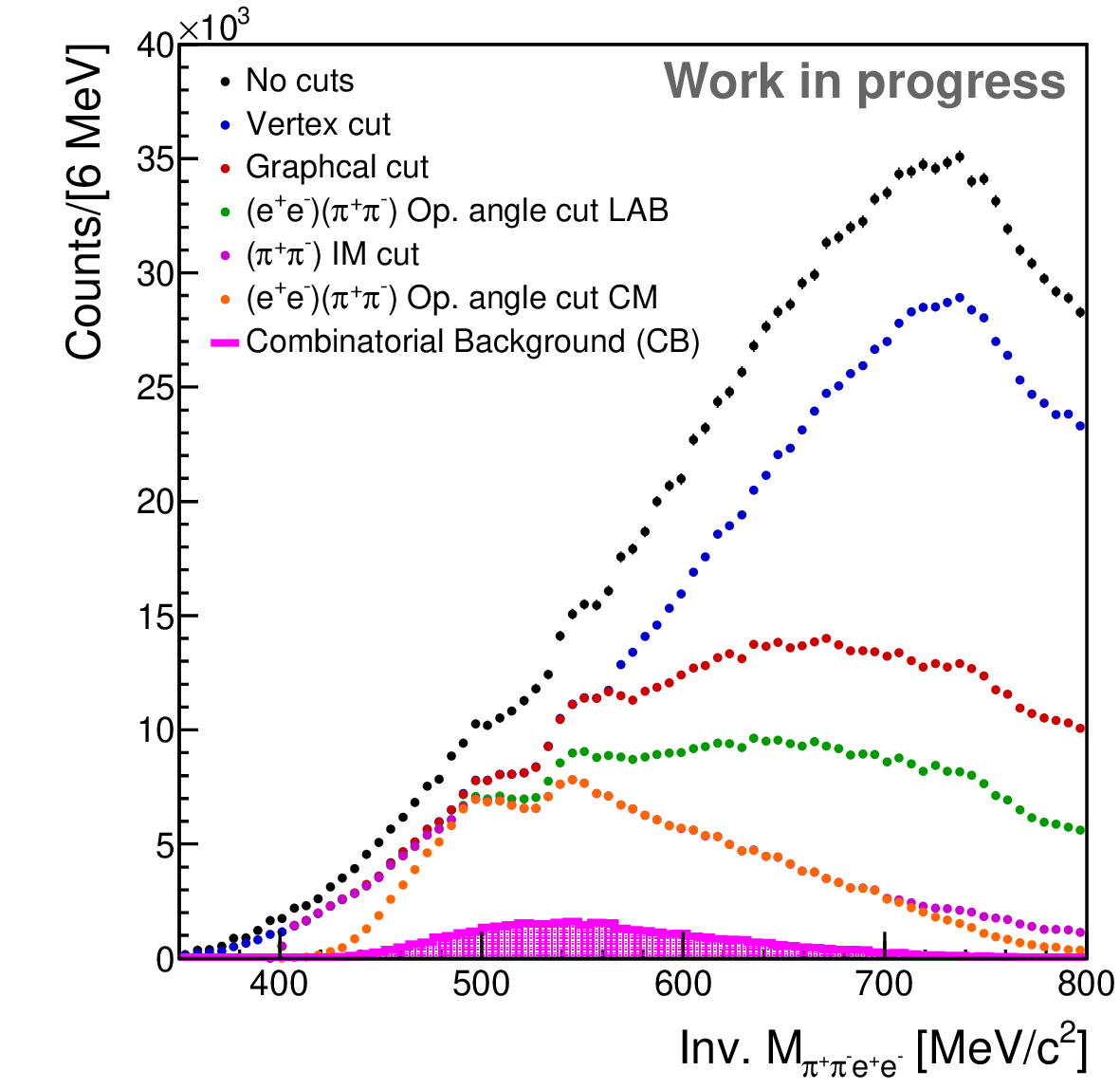}
   \includegraphics[width=6.25cm]{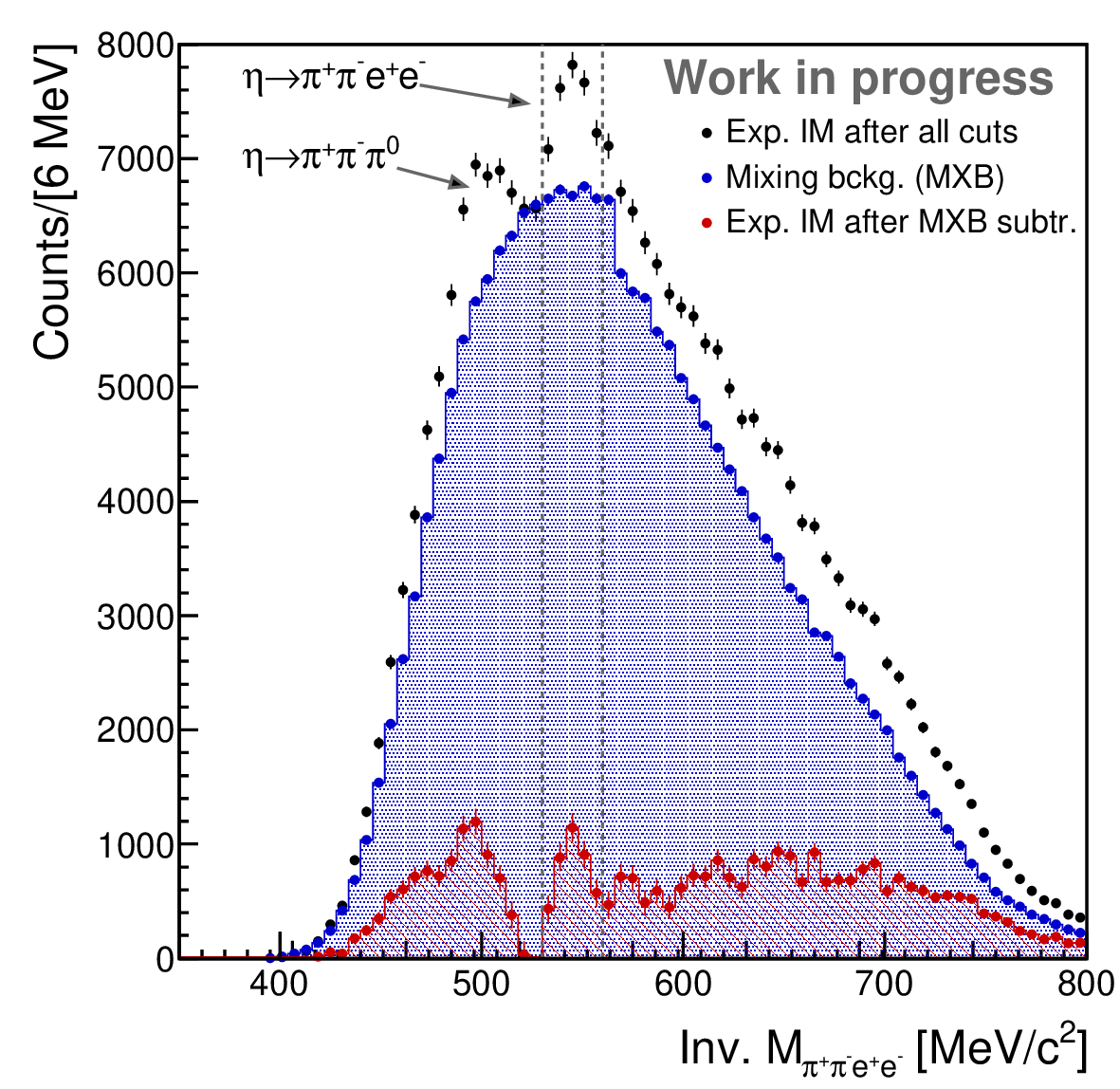}
   }
\caption{Distribution of the invariant mass for the identified $\pi^+\pi^-e^+e^-$ hypothesis. 
{\bf (left)} Result of consecutive kinematical cuts leading to the reduction of the multipion background. 
{\bf (right)} Comparison after all selection cuts: black points represent the experimental distribution after the subtraction of the combinatorial background from like-sign lepton pairs; blue points show the background estimated with the event–mixing method; red points correspond to the difference between the experimental data (black) and the mixed-event background estimation (blue).}

\label{Fig:f3}
\end{figure}

After applying all the cuts described above, the resulting invariant mass spectrum of $\pi^+ \pi^- e^+ e^-$ is obtained (shown in Fig.~\ref{Fig:f3}~(left)). The invariant mass distribution shows a significant reduction in the background, especially for higher masses, and an enhanced signal of the $\eta$ decays. The obtained distribution is still contaminated by combinatorial background events originating from leptons due to the external conversion of photons, as well as from Dalitz decays of single or multiple neutral pions associated with charged pion pairs. Therefore, for further background reduction, the amount of dilepton contamination from combinatorial events has been determined using the formula: $N_{CB} = 2 \sqrt{N_{e^+e^+\pi^+\pi^-} \cdot N_{e^-e^-\pi^+\pi^-}}$, where $N_{e^+e^+\pi^+\pi^-}$ and $N_{e^-e^-\pi^+\pi^-}$ denote the multiplicities of the like-sign lepton pairs~\cite{Gazdzicki:2000fy,NA38:2000wlp,HADES:2011nqx}. 
This procedure removes the $e^{+}e^{-}$ component of combinatorial background while maintaining the two pion background, which is largely correlated.
These correlations are expected from the mechanism of multipion production, which, at these energies, is dominated by double resonance ($N^{*}/\Delta$) production.
The final invariant mass of the  $\pi^+ \pi^- e^+ e^-$ after combinatorial background subtraction is shown in Fig.~\ref{Fig:f3}~(right) as black points.
The obtained invariant mass  distribution reveals  very clearly visible structures originating from: $\eta\rightarrow \pi^0\pi^+\pi^-$ and $\eta\rightarrow \pi^+ \pi^- e^+ e^-$, indicated in Fig.~\ref{Fig:f3}~(right), in accordance with simulations. 

As a first attempt to reproduce the shape of the multipion background, the event–mixing method was applied. In this technique, tracks from different events with similar global properties are combined to build uncorrelated pairs. This provides a reliable description of the smooth background component if the background is truly uncorrelated. The mixed-event spectrum is then normalized to the data at a chosen reference range (520 - 527~MeV). Figure~\ref{Fig:f3}~(right) shows the estimated background obtained with the event–mixing technique, which approximates the experimental background. 
As one can see, this background fails to reproduce the shape and the yield above the $\eta$ peak. It is due to the correlation between pions stemming from resonance decays, as discussed above. However, in the mass region spanned by $\eta$ decays, the mixed events seem to describe the background well.

Therefore, in order to calculate the number of expected $\eta$ mesons, the mixed–event background was used. The number of signal events from $\eta$ decays is calculated as the difference between the total number of events in this region and the area below the estimated mixing background. The calculated number of events amounts to about 5000 in the left peak, corresponding to the $\eta \to \pi^{+}\pi^{-}\pi^{0} (e^+e^-\gamma)$ decay, and about 2750 in the right peak, corresponding to the $\eta \to \pi^{+}\pi^{-}e^{+}e^{-}$ decay. These values represent the estimated number of reconstructed $\eta$ mesons in the respective final states, which will be used to extract upper limits for ALP decays.

\section{Summary}
The analysis of the ALP search in the rare decay channel $\eta \to \pi^+\pi^- e^+ e^-$ with the HADES experiment is currently being developed. Dedicated procedures for particle selection and the suppression of multipion background were presented to enhance the identification of $\eta$ mesons. Within the analyzed data sample, further refinement of the background description is required in order to improve the signal-to-background sensitivity. The strategy discussed here provides the basis for a more precise evaluation of potential ALP signatures. The ongoing work will include the optimization of the selection criteria and validation with extended data samples, with the aim of determining the upper limit for the ALP particle in the 0-200~[MeV] mass range.

\section*{Acknowledgments}
This work was supported by the National Science Centre, Poland (NCN)  under grant No. SONATA-BIS 13 no. 2023/50/E/ST2/00673, and by the "Research support module" as part of the "Excellence Initiative - Research University" program at the Jagiellonian University in Krak\'ow. 

\bibliographystyle{unsrt}
\bibliography{bib2}

\end{document}